\documentstyle[12pt]{article}
\newcommand{\CGC}{\mbox{$C^{\infty}_{c}(G,{\bf C})$}}

\newcommand{\calA}{\mbox{${\cal A}$}}

\newcommand{\za}{\mbox{${\cal Z}({\cal A})$}}

\begin{document}

\title{EMERGENCE OF TIME}

\author{Michael Heller,\thanks{Correspondence address: ul.
Powsta\'nc\'ow Warszawy 13/94, 33-110 Tarn\'ow, Poland.  E-mail:
mheller@wsd.tarnow.pl} \\ Vatican Observatory, \\ V-00120 Vatican
City State
\and
Wies{\l}aw Sasin, \\ Institute of Mathematics, \\ Warsaw University
of Technology, \\ Plac Politechniki 1, 00-661 Warsaw, Poland}
\maketitle

\begin{abstract}
In the groupoid approach to noncommutative quantization of
gravity, gravitational field is quantized in terms of a
$C^*$-algebra \calA \ of complex valued functions on a groupoid
$G=E\times \Gamma $, where $E$ is a suitable space and $\Gamma $
a group of fundamental symmetries. In the noncommutative
quantum gravitational regime the concepts of space and time are
meaningless. We study the ``emergence of time'' in the
transitions process from the noncommutative regime to the
standard space-time geometry. Precise conditions are specified
under which modular groups of the von Neumann algebra generated
by \calA \ can be defined. These groups are interpreted as
representing a state depending time flow of a quantum
gravitational system. If the above conditions are further
refined one obtains a state independent time flow. We show that
quantum gravitational dynamics can be expressed in terms of
modular groups.
\end{abstract}
\vspace{1cm}
\section{Introduction}
The issue of time is one of the most discussed problems in the
research domain known as ``quantum gravity''. Very often the
opinion is expressed that the correct formulation of
questions concerning the time issue could pave the way towards
the looked for theory of quantum gravity (for a sample of
references see \cite{TimeDisc}). In many working models, the
standard (coordinate) time of macroscopic physics either becomes
``imaginary'' (i.e., acquires fully spatial properties)
\cite{H-H}, or entirely loses its meaning (see \cite{TimeDisc}).
Carlo Rovelli, in the series of papers (\cite{R90,R91,Rov}), not
only has propagated the idea that at the fundamental level a
well defined concept of time is totally absent, but also has
obtained some interesting results showing how could physics be
done without the usual notion of time.
\par
In \cite{HSL} we have proposed a scheme for quantizing gravity
in which the quantum gravity regime is modelled by a
noncommutative geometry. In this geometry all local concepts
(such as time instant, point location and their neighbourhoods)
are meaningless. Space-time appears only in the transition
process from the noncommutative regime to the macroscopic
physics. (It is a scheme rather than a full theory since some
its main ingredients, such as the group of fundamental
symmetries, remain unspecified; their choice is left for future
developments). The aim of the present paper is to study the
emergence of time when gradually going from the noncommutative
regime to the usual space-time geometry.  We prove some
mathematically rigorous results responsible for the main stages
of this process. Our results are based on the work by Connes and
Rovelli \cite{ConRov}; it is surprising how the postulates of
these authors, regarding the future theory of quantum gravity,
agree with our scheme.
\par
To make the present paper self-contained, in Section 2, we
briefly summarize our approach to noncommutative quantization of
gravity. The algebra \calA \ determining the noncommutative
geometry, which models the quantum gravity regime, is an algebra
of complex valued functions on a groupoid of fundamental
symmetries with convolution as multiplication. The first
important result is that if we make a suitable identifications
of elements of the algebra \calA \ (if we make a quotient $\calA
/N_{\omega_q}$ where $N_{\omega_q}$ is an ideal of \calA \ and
$\omega_q $ is a state on \calA ), the one-parameter group
$\alpha_t, \, t\in {\bf R}$, of automorphisms of the von Neumann
algebra ${\cal R}$ generated by the algebra \calA , the
so-called {\em modular group\/}, can be defined. This group is
interpreted as a time flow (dependent on the state $\omega_q$).
This is proved in Section 3. It turns out that if we further
identify elements of \calA \ (with the help of the {\em inner
equivalence\/} relation), the modular group can be made state
independent. This is explained and proved in Section 4. In
Section 5, we show that the dynamical equation of our scheme can
be expressed in terms of modular groups (when these groups are
available). Some interpretative comments are given in Section 6.
\par
\section{A noncommutative scheme for quantization of
gravity}
In this Section we briefly summarize our approach to quantum
gravity. It is based on the direct product $G =E\times\Gamma$,
regarded as a smooth groupoid (for definition see
\cite{Renault}), where $E$ is an n-dimensional smooth manifold
(or a structured space of constant dimension, see
\cite{HS}) and $\Gamma$ a Lie group acting on $E$ (to the
right). Elements of $ \Gamma $ are ``fundamental symmetries'' of
our theory (heuristically, we could think of $E$ as of the total
space of the fibre bundle of frames over space-time $M$, and of
$\Gamma $ as of its structural group, i.e., a connected
component of the Lorentz group).  The correct choice of $\Gamma$
is left for the future development of the proposed scheme.
\par
Now, we define the algebra $\calA=\CGC$ of smooth compactly
supported complex-valued functions on $G$ with the convolution
\[
(a*b)(\gamma ):=\int_{G_p}a(\gamma_1)b(\gamma_2),
\]
as multiplication, where $G_p$ is a fiber of $ G$ over $p\in E$,
$a,b\in\calA$, $\gamma =\gamma_1\circ\gamma_2,\, \gamma
,\gamma_1,\gamma_2\in G_p,\;g\in\Gamma$.   \calA\ is also an
involutive algebra with involution defined as $a^{*}(\gamma
)=\overline { a(\gamma^{-1})}$.  
\par
Now, we develop a noncommutative geometry as determined by the
algebra \calA\ in terms of derivations of this algebra (see
\cite{deriv}).
\par
Let Der${\cal A}$ be the set of all derivations of the algebra $
{\cal A}.$ It is a \za-module where \za\ denotes the center of
\calA. Der\calA\ can be thought of as a noncommutative
counterpart of the module of vectors fields. The pair
$(\calA,V)$, where $V$ is a \za-submodule of Der\calA, is called
{\em differential algebra}. In our case $\calA =\CGC$, and as
$V$ we choose those derivations of \calA\ which are naturally
adapted to the structure of $G=E\times\Gamma$ (as a direct
product), i.e., all those $v\in$Der\calA\ which can be presented
in the form $v=v_E+v_{\Gamma}$ where $v_E$ is the "component" of
$v$ parallel to $E$, and $v_{\Gamma}$ the "component" of $ v$
parallel to $\Gamma$.
\par
By a {\em metric\/} on the ${\cal Z}({\cal A} )$-submodule $V$
we understand a \za -bilinear non-degenerate symmetric mapping
$g :V\times V\rightarrow\calA$. We chose the metric of the form
\[
g=pr_E^{*}g_E+pr_{\Gamma}^{*}g_{\Gamma}
\]
where $g_E$ and $g_{\Gamma}$ are metrics on $ E$ and $\Gamma$,
respectively, and $pr_E$ and $pr_{\Gamma }$ are the obvious
projections. This choice of $V$ and $g$ is naturally adapted
to the product structure of $G$ but, if necessary, we could try
other choices as well.
\par
It turns out that now we can define the linear connection
(essentially by using the Koszul formula), curvature and the
Ricci operator ${\bf R}:V\rightarrow V$ (the counterpart of the
Ricci tensor with one index up and one index down). For detail
the reader should consult reference \cite{HSL}.  This allows us
to define the {\em noncommutative Einstein equation\/} in the
operator form
\begin{equation}
{\bf R}+2\Lambda {\bf I}=0\label{R2}
\end{equation}
where $\Lambda$ is a constant related to the usual cosmological
constant, and ${\bf I}$ is the identity operator (the factor 2
appears as the result of our conventions, see \cite{HSL}).  It
should be expected that ``at the fundamental level'' there is
only ``pure noncommutative geometry'', therefore we assume that
at this level there is no  ``matter source'' (such as a
counterpart of the energy-momentum tensor), but for the sake of
generality we keep $\Lambda$ in the equation (if necessary we
can always put $\Lambda =0$).
\par
The set ker${\bf G}:=\{v\in V:{\bf G}(v) =0\}$, where ${\bf
G}={\bf R}+2\Lambda${\bf I,} is a \za-submodule of $V$; it gives
a solution of eq. (\ref{R2}). The differential algebra
$(\calA,{\rm k}{\rm e}{\rm r}{\bf G})$, where $\calA=\CGC$ is
called {\em Einstein algebra\/} (strictly speaking only ker{\bf
G} is determined by eq. (\ref{R2}).
\par
Now, we introduce the representation of the algebra $
{\cal A}$, the so-called {\em Connes representation},
in the Hilbert space ${\cal H}=L^2(G_q)$
\[
\pi_q:\calA\rightarrow {\cal B}({\cal H}),
\]
where ${\cal B}({\cal H})$ denotes an algebra of bounded
operators on ${\cal H}$, with the help of the formula
\begin{equation}
(\pi_q(a)\psi )(\gamma )=\int_{G_q}a(\gamma_1)\psi
(\gamma_1^{-1}\gamma ),\label{Conrepr}
\end{equation}
with $\gamma
=\gamma_1\circ\gamma_2,\;\gamma ,\gamma_ 1,\gamma_2\in
G_q, \,q\in E\;\psi\in {\cal H},\;a\in {\cal A}$ (see
\cite[p. 102]{Connes}). It can be shown that the
completion of \calA\ with respect to the norm
\[
\parallel a\parallel\,={\rm s}{\rm u}{\rm p}_{
q\in E}\parallel\pi_q(a)\parallel
\]
is a $C^{*}$-algebra (let us notice that $q$ can formally be
understood as the pair $(q,e)\in G$ where $e$ is the unit of
$\Gamma $). We shall denote this algebra by $ {\cal E}$ and call
{\em Einstein} $C^{*}${\em -algebra}.
\par
The next natural step is to perform quantization with the help
the usual C$^{*}$-algebraic method (see, for instance,
\cite{Thirring}).  A quantum gravitational system is represented
by an Einstein $C^{*}$-algebra ${\cal E}$, and its observables
by Hermitian elements of ${\cal E}$.  If $a$ is a Hermitian
element of ${\cal E}$, and $\phi $ a state on ${\cal E}$ then
$\phi (a)$  is the expectation value of the observable $ a$ when
the system is in the state $\phi$.  The essentially new
ingredient of our approach is the postulate according to which
the dynamical equation of a quantum gravitational system is
\begin{equation}
i\hbar\pi_q(v(a))=[\pi_q(a),F]\label{dyneq}
\end{equation} 
for every $q\in\ E,\;\psi\in L^2(G_q)$. Here $ v\in {\rm k}{\rm
e}{\rm r}{\bf G}$ and in this way generalized Einstein's
equation (\ref{R2}) is coupled to quantum dynamical equation
(\ref{dyneq}). $F$ is a Fredholm operator, i.e. an operator $
F:{\cal H}\rightarrow {\cal H}$ such that $F({\cal H})$ is
closed and the dimensions of its kernel and cokernel are finite.
The Planck constant $\hbar $ should be regarded as measuring a
deformation from commutativity. Eq. (\ref{dyneq}) is a
noncommutative counterpart of the Schr\"odinger equation in the
Heisenberg picture of the usual quantum mechanics{\bf .}
Equation (\ref{dyneq}) acts on that Hilbert space $ L^2(G_q)$
which should be regarded as a counterpart of the Hilbert space
in the position representation in quantum mechanics.
Accordingly, the quantity $|\psi (\gamma )|^2$ is the
probability density of the ``fundamental symmetry'' $\gamma\in
G_q$ to occur.
\par
In \cite{HSL}\ it has been shown that the above sketched gravity
quantization scheme correctly reproduces the usual general
relativity (on space-time) and quantum mechanics (in the
Heisenberg picture) when the algebra Der${\cal A}$ is suitably
restricted to its center ${\cal Z}({\cal A})$ (or to some subset
of $ {\cal Z}({\cal A})$).
\par

\section{State dependent flow of time}
We shall assume that $\calA = C_c^{\infty}(G,{\bf C})$ is
already suitably completed to form a C$^{*}$-algebra. Let us
consider a state $\omega :{\cal A}\rightarrow {\bf C}$ on ${\cal
A}$, i.e., a positive, linear and normed functional on ${\cal
A}$. From the Gelfand-Naimark-Segal theorem (see \cite{Murphy})
it follows that there exists exactly one representation $\pi_{
\omega}:{\cal A}\rightarrow {\rm E}{\rm n}{\rm d}
{\cal H}_{\omega}$ of the algebra ${\cal A}$ on a Hilbert space
${\cal H}_{\omega}$, and the vector $
\xi_{\omega} \in {\cal H}_\omega $ such that
\begin{enumerate}
\item[(i)]
Lin$(\pi_{\omega}({\cal A})\xi_{\omega })={\cal H}_{
\omega},$
\item[(ii)]
$\omega (a)=(\pi_{\omega}(a)\xi_{\omega},\xi_{
\omega})$
\end{enumerate}
for every $a\in {\cal A}$.
\par
Let ${\cal R}$ be a von Neumann algebra generated by $
\pi_{\omega}({\cal A})$, i.e., ${\cal R}=(\pi_\omega ({\cal
A}))^{ \prime\prime}$, where ${\cal A}'$ denotes the commutant
of ${\cal A}$. We shall consider a one-parameter group of
automorphisms $\alpha_ t:{\cal R}\rightarrow {\cal R},\,t\in
{\bf R}$. Let the vector $
\xi_{\omega}$ be a cyclic and separating vector in the Hilbert
space $ {\cal H}_{\omega}$ $(\xi$ is {\em separating\/} in
${\cal A}$ if it is cyclic in $ {\cal A}'$). We define the
operator  $S:\,{\cal R}\rightarrow {\cal R}$ by
\[
S(b)(\xi_{\omega})=b^{*}(\xi_{\omega})
\]
for $b\in {\cal R}$. It can be shown \cite[p. 43]{Connes} that
$ S$ has the following properties
\begin{enumerate}
\item[(i)]
$S=S^{-1}$,
\item[(ii)]
the operator $J=S|S|^{-1}$ satisfies the condition $ J{\cal
R}J^{-1}={\cal R}',$
\item[(iii)]
the operator $\Delta =|S|^2=S^{*}S$ satisfies the condition $
\Delta^{it}{\cal R}\Delta^{-it}={\cal R}$ for every $
t\in {\bf R}.$
\end{enumerate}
From these properties it follows that
\[
S=J\cdot\Delta^{1/2}
\]
where $J$ is an antiunitary operator, and $\Delta$ a
self-adjoint, positive operator. The Tomita-Takesaki theorem 
\cite{TT} asserts that the mappings $\alpha_t:\,{\cal
R}\rightarrow {\cal R} ,\;t\in {\bf R}$, given by
\begin{equation}
\alpha_t(b)=\Delta^{-it}b\Delta^{it},\label{eq1}
\end{equation}
$b\in {\cal R}$, form a one-parameter group of automorphisms of
the von Neumann algebra ${\cal R}$. It is called the {\em
modular group of modular automorphisms\/} of the state $\omega$
on the von Neumann algebra $ {\cal R}$, or the {\em modular
group\/} for brevity. Connes and Rovelli \cite{ConRov} have
interpreted this one-parameter group as a state dependent time
in the framework of noncommutative geometry, provided that the
state $\omega $ is of the form $\omega (a)={\rm T}{\rm
r}[a\omega ]$, for every $a\in {\cal A}$, where ${\cal A}$ is
any C$^{ *}$-algebra of bounded linear operators on a Hilbert
space. The last assumption was necessary in order to connect the
state dependent time flow with statistical (thermodynamical)
properties of the considered system (see Sec. 6 below). In the
following, we shall interpret $\alpha_t$ in terms of time, but
we shall remain strictly within the algebraic approach.
\par
The question arises: what is the relationship between the GNS
representation of ${\cal A}$ and the Connes representation of
${\cal A}$ as far as time properties are concerned? The answer
to this question is given by the following theorem:
\par
{\em Theorem 1:\/} Let $\pi_q:\,{\cal A}
\rightarrow {\rm E}{\rm n}{\rm d}{\cal H}$ be the Connes
representation of the algebra ${\cal A}=C_c^{\infty}(G,{\bf C})$
in the Hilbert space $ {\cal H}=L^2(G_q)$ in which there is a
cyclic vector $\xi_0$.  There exists the unique state
$\omega_q=(\pi_q(a)\xi_0,\xi_0)$, for every $a\in {\cal A}$, and
\[
\pi_{\omega_q}(a)[b]=[\pi_q(a)(b)]
\]
is the GNS representation of ${\cal A}$. Here [...] denotes an
element of the quotient space ${\cal A}/N_{\omega_q}$,
$N_{\omega_ q}$ being the ideal $N_{\omega_q}=\{a\in {\cal A}
:\,\omega_q(aa^{*})=0\}$ of the algebra ${\cal A}$.
\par
{\em Proof:\/} For a cyclic vector $\xi_
0\in {\cal H}=L^2(G_q)$, we define the state
$\omega_q(a)=(\pi_q(a)\xi_0,\xi_0)=(a*\xi_0,\xi_ 0),\,a\in {\cal
A}$. Now, $N_{\omega_q}=\{a\in {\cal A}:\omega_q(aa^{*})
=0\}=\{a\in {\cal A}:(\pi_q(aa^{*})\xi_0,\xi_ 0)=0\}.$ A suitable
completion of ${\cal A}/N_{\omega_q}$ gives us the Hilbert space
${\cal H}_{\omega_q}$. We define the GNS representation
$\pi_{\omega_q}:\,{\cal A}\rightarrow {\rm E} {\rm n}{\rm
d}{\cal H}_{\omega_q}$ by
\[
(\pi_{\omega_q}(a))[b]=[a*b]=[\pi_{\omega_q}(a)(b)]
\]
with $a\in {\cal A},$$\,[b],[a*b]\in {\cal H}_{
\omega_q}$. $\Box$
\par
It is now evident that we can construct the modular group $
\alpha_t$ of the state $\omega_q$ on the van Neumann algebra 
${\cal R}=(\pi_{\omega_q}({\cal A}))^{\prime\prime}$. If we
interpret this one-parameter group as a time flow, we have the
interesting conclusion: To have a (state dependent) time flow in
the noncommutative space, determined by the algebra $ {\cal
A}=C_c^{\infty}(G,{\bf C})$, we must form the quotient space
${\cal A}/N_{\omega_ q}$, i.e., we must glue together some
elements of ${\cal A}$. In other words, in the original
noncommutative regime (as determined by ${\cal A}$), in
principle, there is no time; a state dependent time flow emerges
only in the process of a suitable coarse graining of the
original space (i.e., in the process of forming the quotient
${\cal A}/N_{\omega_q}$).
\par

\section{State independent time flow}
Let us consider the subset
\[
{\cal U}=\{u\in {\cal A}:\;uu^{*}=u^{*}u={\bf I}\};
\]
of the algebra \calA ; it forms a group, called the {\em unitary
group\/} of $ {\cal A}$.
\par
{\em Lemma\/} 2{\em :\/} Let ${\cal U}$ be the unitary group of
$ {\cal A}=C_c^{\infty}(G,{\bf C})$ and let $u\in {\cal A}$
implies $u_q\in L^2(G_q)$, then the Connes representation
$\pi_q:\,{\cal U}\rightarrow {\rm E}{\rm n}{\rm d} {\cal
H},\;{\cal H}=L^2(G_q)$, given by
\[
\pi_q(u)\xi =u_q*\xi ,
\]
for $u\in {\cal U},\,\xi\in {\cal H}$, is a unitary
representation.
\par
{\em Proof:\/} A representation $\pi$ is unitary if it
preserves scalar products, or equivalently if: $\pi (u)^{*}=\pi
(u^{-1}) =\pi (u)^{-1}$, for $u\in {\cal U}$. Let $\xi ,\eta\in
{\cal H},\,u\in U$. One has
\[
(\pi_q(u)\xi ,\pi_q(u)\eta )=(u_q*\xi ,u_q*
\eta )=
\] \[
=\int_{G_q}(u_q*\xi )^{*}(u_q*\eta )=
\] \[
= \xi^{*}*u_q^{*}*u_q*\eta =(
\xi ,\eta ).\Box
\]
\par 
Since the Connes representation of ${\cal U}$ is unitary
it is also irreducible (see \cite[p. 127]{Kirillov}) and,
consequently, there exists a one-to-one correspondence between
such representations and pure states on $ {\cal A}$.
\par
Let ${\cal R}$ be a von Neumann algebra. An automorphism $\alpha
:\,{\cal R}\rightarrow {\cal R}$ is called {\em inner\/} if
there exists a nontrivial element $ u$ of the unitary group
${\cal U}$ (i.e., $u\neq {\bf I}$) such that
\[
\alpha (a)=u^{*}au
\]
for every $a\in {\cal R}$. Let us consider two automorphisms $
\alpha'$ and $\alpha^{\prime\prime}$ of ${\cal R}$. 
These automorphisms are said to be {\em inner equivalent,} $
\alpha'\sim\alpha^{\prime\prime}${\em ,\/}  if 
there is an inner automorphism $\alpha_{inner}$ such that
\[
\alpha^{\prime\prime}=\alpha_{inner}\alpha'
\]
or, equivalently,
\[
u\alpha^{\prime\prime}(a)=\alpha'(a)u
\]
for every $a\in {\cal R}$ and some $u\in {\cal U}$. In other
words, two automorphisms of ${\cal R}$ are inner equivalent if
they differ by an inner automorphism. The set of all equivalence
classes of this relation is called a group of {\em outer
automorphisms\/} of ${\cal R},$ denoted by $ {\rm O}{\rm u}{\rm
t}({\cal R})$ .
\par
In general, the modular group $\alpha_ t$ of a state $\omega$ on
the von Neumann algebra ${\cal R}$ does not consists of inner
automorphisms, but it projects down to a nontrivial
one-parameter group $\tilde{
\alpha}_t,\,t\in {\bf R}$, in ${\rm O}{\rm u}{\rm t}
({\cal R})$. According to the cocycle Radon-Nikodym theorem (see
\cite[p. 44]{Connes}) two modular automorphisms $\omega_1$ and
$\omega_ 2$ on ${\cal R}$ are inner equivalent.  Consequently,
the one-parameter group $\tilde{\alpha }_t$ does not depend of
the choice of $\omega$ on ${\cal R}$, and $\tilde{
\alpha_t}$ can be interpreted as a state independent,
``canonical time'' in the considered noncommutative space (see
\cite{ConRov}).
\par
Lemma 2 allows us to apply the above construction to our case.
We thus obtain a one-parameter group in ${\rm O} {\rm u}{\rm
t}({\cal R})$, where ${\cal R}=(\pi_{
\omega_q}({\cal A}))^{\prime\prime}$ (see remarks following
theorem 1), which is independent of the choice of the state
$\omega_q$.  This can be interpreted in the following way. If we
make a suitable ``coarse graining'' of our state dependent time
$\alpha_t$ (i.e. if we change from ${\cal R}$ to Out$({\cal
R})$), we obtain a ``time flow'' in our noncommutative space
which depends only on the algebra ${\cal A}=C_c^{\infty}(G,{\bf
C} )$.
\par

\section{Noncommutative dynamics}
An interesting feature of our scheme for quantizing gravity is
the fact that although in the noncommutative regime there is no
time (in the usual sense), dynamics can be done in terms of
derivations of the algebra ${\cal A}=C_c^{\infty}(G,{\bf C})$
(see eq.  (\ref{dyneq})). If the modular groups of Section 3 are
interpreted as a sort of (state dependent) time, the question
arises whether these groups can be related to the noncommutative
dynamics.  To answer this question is the aim of the present
section.
\par
From eq. (\ref{eq1}), defining modular groups, we have
\[
\alpha_t(a)=e^{-it\ln\Delta}ae^{it\ln\Delta}.
\]
By differentiating this expression we obtain
\[
\frac d{dt}\alpha_t(a)|_{t=0}=i[a,\ln\Delta ].
\]
\par
On the other hand, let us notice that the modular group $
\alpha_t$ determines the derivation of the von Neumann algebra $ {\cal
R}$. Indeed, if $v\in$ Der${\cal R}$, and the system is in a
state $\omega ,$ we have
\[
v(\pi_{\omega}(a))=\frac d{dt}|_{t=0}(\alpha_
t(\pi_{\omega}(a))=
\]
\begin{equation}
=i[\pi_{\omega}(a),\ln\Delta 
]=i{\rm a}{\rm d}_{\ln\Delta}(\pi_{\omega}(a)
).\label{eq2}
\end{equation}
\par
When we change from the Connes representation $
\pi_q$ of the algebra ${\cal A}$ 
to its CNS representation $\pi_{\omega_q}$ [by forming the
quotient $ {\cal A}/N_{\omega_q}$ (see theorem 1)] the algebra
becomes more coarse, and one-parameter modular groups emerge.
Let us assume that both the derivation $v$ and the Fredholm
operator $F$, appearing in eq.  (\ref{dyneq}), are invariant
with respect to the equivalence relation $
\sim$ defined by: $a\sim b$ 
iff $a-b\in N_{\omega_q}$, for every $a,b\in\pi_ q({\cal A})$;
i.e., we assume that $a\sim b$ implies $F(a)\sim F(b)$, for
every $a,b\in\pi_q({\cal A} )$, and similarly for derivation
$v$.  Then eq. (\ref{dyneq}) takes the form
\[
i\hbar\bar {v}([\pi_q(a)]=[[\pi_q(a)],\bar{F}]
\]
where $\bar {v}([a])=[v(a)]$, $\bar {F}([a])= [F(a)]$, or
taking into account (\ref{eq2}),
\[
\hbar\frac d{dt}|_{t=0}(\alpha_t(\pi_{\omega_
q}(a)))=[\pi_{\omega_q}(a),\bar {F}]
\]
where $\bar{F}=\ln\Delta$.  We can see that if the ``modular
time'' $\alpha_t$ is available, it can serve as a dynamical
time. 
\par

\section{Interpretation}
To define the modular group $\alpha_t$, Connes and Rovelli
\cite{ConRov} have distinguished the state on \calA \ of the form
$\omega (a)= {\rm Tr}[a\omega]$ for every $a \in \calA $ (which,
in the language used by physicists is a density matrix). Owing
to this choice they were able to argue that the time flow has a
statistical (thermodynamic) origin. They emphasize that it is
not only the arrow of time that emerges in this way, but also
the time flow itself. Indeed, the exegesis of the mathematical
formalism presented in the preceding sections allows us to
interpret the one-parameter groups $\alpha_t$ and
$\tilde{\alpha_t}$ as a state dependent and state independent
time flows, respectively (possibly even without a sharply
determined arrow of time). The non-local character of the
algebra $\calA = C_c^{\infty}(G,{\bf C})$ suggests that in the
noncommutative regime there is no multiplicity (there are no
individuals), and consequently there can be neither statistics
(in the usual sense) nor thermodynamics. Only after suitable
identifications of elements of ${\cal A}$ have been made (with
the help of the corresponding equivalence relations) the groups
$\alpha_t$ and $\tilde{\alpha}_t$ appear, and one can speak of a
certain temporal succession. It is not that time has its origin
in thermodynamics, but rather both thermodynamics and time have
their origin in a transition from the noncommutative geometry as
determined by the full algebra \calA \ to a geometry in which
the groups $\alpha_1$ and $\tilde{\alpha }_1$ are meaningful.
\par
The fact that modular groups lead to correct classical limits
remains in agreement with this interpretation. For instance, as
shown by Connes and Rovelli \cite{ConRov}, the modular group of
the Gibbs state gives the standard time flow modulo the choice
of the time unit, and in the case of classical mechanics it goes
onto the usual Hamiltonian evolution.
\par
Moreover, we could even claim that the probabilistic nature of
the standard quantum mechanics has its origin in the ``phase
transition'' from the noncommutative regime to the usual
physics. For what does it mean that if $a$ is a Hermitian
element of the Einstein algebra ${\cal E}$, and $\phi $ a state
on ${\cal E}$, then $\phi (a)$ is the expectation value of the
observable $a$ when the system is in the state $\phi $ (see Sec.
2 above)? In the noncommutative regime the terms such as
``expectation value'' seem to be meaningless, and $\phi (a)$ is
just a value of the functional $\phi $ at $a$. The above phrase
could mean only that the expression $\phi (a)$ (and similar
expressions as well) acquires its quantum-probabilistic
interpretation if we go from the noncommutative regime to the
standard quantum mechanics (as described in \cite{HSL}).
\par
It is worth mentioning that Rovelli \cite{R90,R91} has
demonstrated that there exists a natural extension of canonical
Heisenberg-picture of quantum mechanics which remains valid even
if time is not defined. This is exactly the case in our scheme
in which the quantum sector of noncommutative gravity is such an
extension of the Heisenberg picture. The usual time evolution
appears only within a ``non-quantum gravity approximation''.
\par
\vspace{0.5cm}
\noindent
ACKNOWLEDGMENT: We thank Professor Piotr Hajac for turning our
attention to paper \cite{ConRov}.

\end{document}